\documentclass[aip,rsi,reprint,graphicx]{revtex4-1} 

\usepackage{amsmath}
\usepackage[dvips]{graphicx}
\usepackage{bm}

\usepackage{graphicx}   
\usepackage{verbatim}   
\usepackage{color}      
\usepackage{subfigure}  
\usepackage{hyperref}   

\usepackage{epstopdf}
\DeclareBoldMathCommand{\bfmu}{\mu}
\newcommand{\be}{\begin{equation}}
\newcommand{\ee}{\end{equation}}
\def\mum{\rm\mu m}

\def\insitu{{\it in~situ\/}}

\begin{document}

\title{Image analysis as an improved melting criterion in laser-heated diamond anvil cell}
\author{Ran Salem} \email{ransl@nrcn.org.il}
\affiliation{Physics Department, Nuclear research center Negev, P.O. Box 9001,
Beer-Sheva 84190, Israel}
\author{Shlomi Matityahu}
\affiliation{Physics Department, Nuclear research center Negev, P.O. Box 9001,
Beer-Sheva 84190, Israel}
\affiliation{Physics Department,
Ben-Gurion University of the Negev, Beer Sheva 84105, Israel}
\author{Aviva Melchior}
\affiliation{Physics Department, Nuclear research center Negev, P.O. Box 9001,
Beer-Sheva 84190, Israel}
\author{Mark Nikolaevsky}
\affiliation{Physics Department, Nuclear research center Negev, P.O. Box 9001,
Beer-Sheva 84190, Israel}
\affiliation{School of Physics and Astronomy, Tel Aviv University, 69978,
Tel-Aviv, Israel}
\author{Ori Noked}

\affiliation{Physics Department, Nuclear research center Negev,
P.O. Box 9001, Beer-Sheva 84190, Israel}
\affiliation{Lyman Laboratory of Physics
Harvard University Cambridge, Massachusetts 02138, USA}
\author{Eran Sterer}
\affiliation{Physics Department, Nuclear research center Negev, P.O. Box 9001,
Beer-Sheva 84190, Israel}
\affiliation{Lyman Laboratory of Physics
Harvard University Cambridge, Massachusetts 02138, USA}
\begin{abstract}
The precision of melting curve measurements using laser-heated
diamond anvil cell (LHDAC) is largely limited by the correct and
reliable determination of the onset of melting. We present a novel
image analysis of speckle interference patterns in the LHDAC as a
way to define quantitative measures which enable an objective
determination of the melting transition. Combined with our
low-temperature customized IR pyrometer, designed for measurements
down to 500K, our setup allows studying the melting curve of
materials with low melting temperatures, with relatively high
precision. As an application, the melting curve of Te was measured
up to~$\rm35\,GPa$. The results are found to be in good agreement
with previous data obtained at pressures up to~$\rm10\,GPa$.
\end{abstract}

\maketitle

\section{Introduction}

The laser-heated diamond anvil cell (LHDAC) has been widely used
to synthesize new materials and to study matter under extreme
conditions of pressure and
temperature.\cite{boehler2000laser,salamat2014situ} One of the
main applications of the LHDAC is the study of high pressure
melting curves of elements and
compounds.\cite{salamat2014situ,Errandonea20062017} Major advances
in this technique in the last three decades have provided a
valuable information on the melting curve of a wide variety of
materials at pressures up to a few megabars and temperatures up to
several thousands of degrees Kelvin. Special attention has been
paid to the improvement of heating
capability,\cite{boehler1990melting,funamori2006heating,goncharov2009laser}
temperature
measurement,\cite{kavner2008precise,campbell2008measurement,du2013mapping}
temperature control,\cite{heinz1991laser} minimization of
temperature gradients within the sampled
region,\cite{panero2001effect,kiefer2005finite} and the correct
identification of the onset of
melting.\cite{jeanloz1996melting,benedetti2008integrated}

Despite the advances described above, many investigations of
apparently simple materials have generated controversy due to
disagreements among
experiments,\cite{dewaele2010high,nguyen2004melting} and between
experiments and
theory.\cite{ANIE:ANIE201308039,belonoshko2006xenon} Some studies
have identified possible errors in temperature
measurement.\cite{deemyad2008temperature} Others have found that
chemical reactions or diffusion can alter the composition of the
sample.\cite{dewaele2010high} In particular, the correct
determination of the melting transition has been a highly
controversial subject.\cite{jeanloz1996melting,geballe2012origin}
A variety of melting criteria have been used, including visual
observation of fluid motion in a speckle interference
pattern,\cite{boehler1993temperatures,errandonea2001systematics}
apearence of diffuse scattering via {\insitu} X-ray
diffraction,\cite{dewaele2007melting,dewaele2010high,anzellini2013melting}
appearance of plateaus in curves of temperature as a function of
laser power,\cite{boehler1990melting,deemyad2005pulsed} formation
of glass or a change of texture upon
quenching\cite{du2013mapping,heinz1987measurement,errandonea2001systematics,ruiz2010microscopic,yang2012flash}
and changes in sample properties such as reflectivity,
absorption\cite{zerr1994constraints,saxena1994temperatures,shen1995measurement}
and resistivity.\cite{schaeffer2012high}  Among these, the first
two are the most commonly used criteria. However, the use of X-ray
diffraction in LHDAC experiments is limited to large-scale
synchrotron radiation facilities. Most small-scale laboratories
therefore employ the direct fluid motion observation as the main
melting criterion. It has been argued, however, that this
criterion has two shortcomings. First, the precise identification
of the onset of fluid motion can be difficult, resulting in an
overestimated melting temperature.\cite{jeanloz1996melting} This
becomes more pronounced as the pressure increases and fluid motion
becomes more sluggish.\cite{boehler2007properties} It is thus
important to define this criterion more objectively, based on
quantitative measures. Second, this criterion becomes increasingly
subjective at temperatures above~$\rm2500\,-\,3000\,K$, as intense
thermal radiation makes it impractical to obtain a reliable image
with adequate contrast across the hotspot within the
sample.\cite{jeanloz1996melting} This limits the validity of this
criterion to relatively low temperatures
(~$T<\rm2500\,-\,3000\,K$).

In this paper we describe an image analysis procedure used to
extract quantitative information on the speckle interference
pattern. This allows us to define several quantitative measures,
the abrupt change of which indicates the onset of melting. To
further corroborate the correct melting point determination, we
have also monitored the temperature as a function of laser power
and interpreted plateaus or change of slope as the onset of
melting. This method is applied to the measurement of the melting
curve of Tellurium (Te) up to~$\rm35\,GPa$. At low pressures, up
to~$\rm10\,GPa$, our measurements are in excellent agreement with
a previous measurement of the melting curve of Te using
large-volume press.\cite{brazhkin1997high} Furthermore, at higher
pressures our measurements are in line with the solid-solid
coexistence curve reported in Ref.\ \onlinecite{hejny2006phase}

The paper is organized as follows: In Sec.\ \ref{Sec1} we describe
the experimental setup including the sample preparation procedure
(Sec.\ \ref{Sec1A}), the optical setup and temperature measurement
(Sec.\ \ref{Sec1B}), and the image analysis of the speckle
interference pattern (Sec.\ \ref{Sec1C}). The measurement of the
melting curve of Te using this improved melting criterion is
presented in Sec.\ \ref{Sec2}. Finally, we summarize and discuss
our results and their implications in Sec.\ \ref{Sec3}.


\section{Experimental}
\label{Sec1}
\subsection{Sample preparation}
\label{Sec1A} Samples of high purity Te~($\rm99.999\%$) were used
without further treatment. The samples were cut and pressed
between diamond culets to form pieces of~$\rm30\,\mum$ thickness
and a diameter of about~$\rm80\,\mum$. The samples were mounted in
a DAC with~$\rm300\,\mum$ diameter culets. Rhenium gaskets, pre
indented to~$\rm50\,\mum$ thickness were used. Thermal isolation
was achieved by loading a thermally isolated epoxy in the laser
drilled $\rm200\,\mum$ diameter sample chamber and drilling
{\insitu} with a~$\rm100\,\mum$ diameter drill bit. The bottom
culet was covered with a~$\rm5\,-\,10\,\mum$ layer of ruby by
crushing a large ruby grain between the bare culets. In some
loads~$\rm3\,-\,5\,\mum$ thick KCL plates were used as an
additional thermal isolation layer. The sample was topped with a
few~$\rm10\,\mum$ ruby chips and the top diamond was mounted on
it. Finally, the cell was cryogenically loaded with~$\rm99.999\%$
argon as a pressurization medium.

\subsection{The optical system and temperature measurement}
\label{Sec1B} 
\begin{figure}[ht]%
\includegraphics[width=\columnwidth]{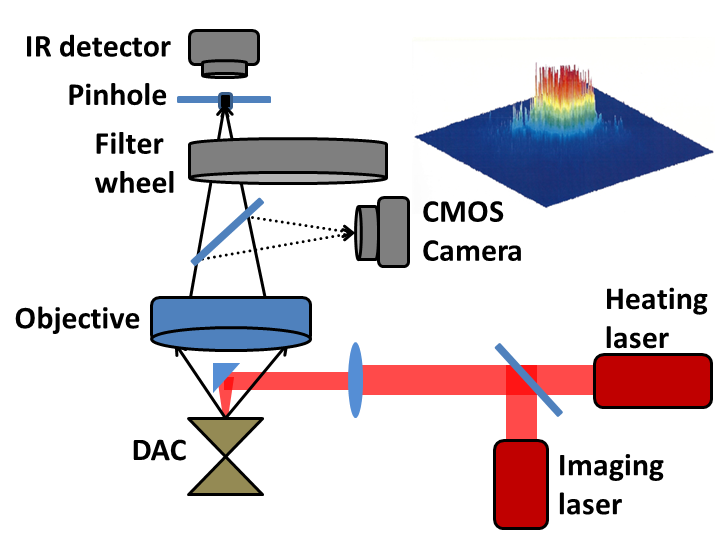}%
\caption{(Color online) Schematic of the optical setup. CW ND:YAG
laser is focused on the sample. Thermal radiation emitted by the
hotspot is focused on the IR detector via a Schwarzschild
objective. An auxiliary HeNe laser is used to generate a speckle
interference pattern on a CMOS camera. An example for a speckle
image obtained by the CMOS camera is shown in the upper right side
of the image.} \label{fig_Opt_Meas}
\end{figure}

The optical setup is shown schematically in Fig.\
\ref{fig_Opt_Meas}; see Ref.\ \onlinecite{shuker2008ir} for more
details. CW Nd:YAG laser is focused to form a hotspot
$\rm50\,\mum$ in diameter at the center of the sample. A homemade
IR pyrometer, designed for measurements down to
500K,\cite{shuker2008ir} is used to measure the hotspot peak
temperature. Thermal radiation emitted by the hotspot is collected
from a small area ($\rm23\,\mum$ in diameter) at the center of the
hotspot into a GaAs photodiode (ThorLabs d10m). In order to ensure
thermal equilibrium conditions, the thermal radiation is
collected $\rm2$ seconds after exposing the sample to the heating
laser beam. The thermal radiation intensity is measured at $\rm8$
different wavelengths in the spectral range
~$\rm1.2\,-\,2.6\,\mum$ using narrow band filters.

The temperature is calculated by fitting the measured spectrum to
Planck's spectral radiance law
\begin{align}
\label{eq:1}&I(\lambda,T)=\varepsilon(\lambda,T)\frac{2\pi
hc^{2}}{\lambda^{5}}\frac{1}{e^{hc/\lambda k^{}_{B}T}-1},
\end{align}
where $\lambda$ is the wavelength, $T$ is the temperature,
$\varepsilon(\lambda,T)$ is the sample emissivity and $h$,
$k^{}_{B}$ and $c$ are Planck's constant, Boltzman's constant and
the speed of light, respectively. Invoking the grey body
approximation which assumes $\varepsilon(\lambda,T)=\varepsilon$
to be a constant independent of wavelength and temperature, we fit
the measured spectrum to Eq.\ \eqref{eq:1} using two free
parameters, the constant emissivity and the temperature. At fixed
pressure and hotspot within the sample, temperatures are measured
with increasing heating beam power.

It should be noted that the emissivity extracted in this fitting
procedure is an effective emissivity, which is strongly affected
by the surface texture and local optical properties of the
hotspot. At the melting transition the surface texture changes and
the effective emissivity can vary and generate errors in the
temperature measurement. For better accuracy of the temperature
measurement we perform a second temperature calculation after
completing a series of temperature measurements at the same
hotspot; each temperature in the series is recalculated assuming a
common emissivity, which serves as a single free parameter for a
series of measurements at the given hotspot. Having determined the
melting point (see details in the next section) for each series of
measurements, we extract the value of the emissivity by fitting
the measured spectra to Eq.\ \eqref{eq:1} for temperatures below
the melting point. The extracted solid phase emissivity is then
used to calculate the temperatures of all the measurements in the
series. Temperatures measured above the melting point might have
larger error due to this procedure. However, the melting
temperature is determined by measurements carried out just before
melting and in the vicinity of the melting transition. Therefore,
errors in temperature measurements above the melting point do not
significantly affect the correct determination of the melting
temperature.

\begin{figure}[ht]%
\includegraphics[width=\columnwidth]{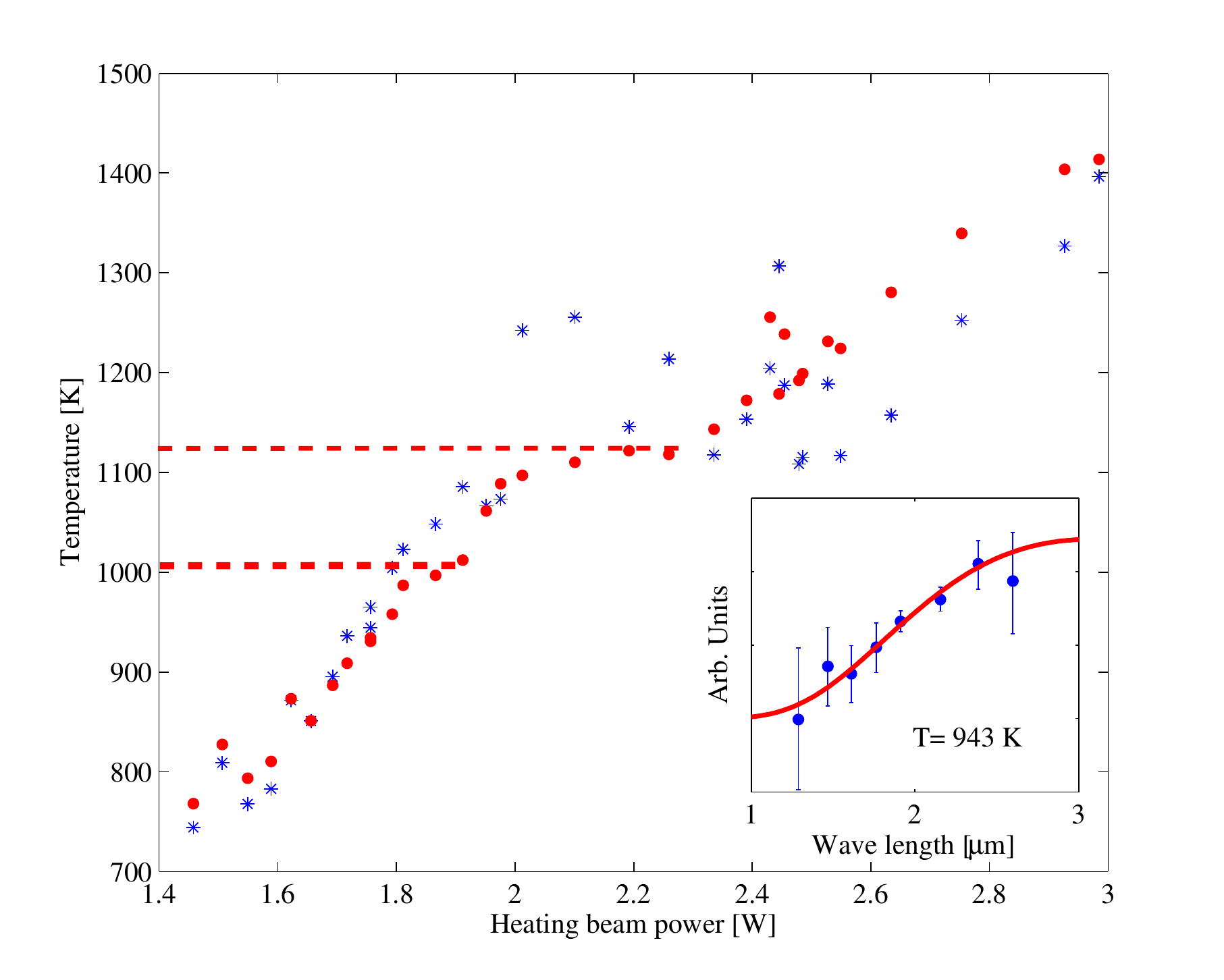}%
\caption{(Color online) Calculated temperatures as a function of
heating beam power. Measurements were performed on Te sample
at~$\rm10.5\,GPa$. As a first approximation the temperature of
each measurement is calculated by fitting the measured spectrum
(inset) to a grey body spectral radiance formula (blue diamonds).
The temperature is then recalculated by fitting the series of
measurements taken at the same position subject to the constraint
of a common emissivity for the entire series (red
points). The dotted red lines correspond to the
melting temperatures of Te ($T^{}_{m}\approx 1000$K) and Ar
($T^{}_{m}\approx 1120$K).}
\label{fig_Temp_Meas}%
\end{figure}

Figure \ref{fig_Temp_Meas} shows a series of temperature
measurements, performed at a pressure of~$\rm10.5\,GPa$, as a
function of heating beam power. As can be observed, temperatures
calculated using two free parameters for each measurement
separately (blue stars) have strong fluctuations around the
melting temperature, $T^{}_{m}\approx\rm1000\,K$. Once the melting
point was determined and the temperatures were recalculated using
a common emissivity (red points in Fig. \ref{fig_Temp_Meas}), one
can observe a plateau around the melting temperature of Te. As
evident from Fig.\ \ref{fig_Temp_Meas}, temperatures below the
melting point are not very sensitive to the fitting procedure (the
corrections are smaller than~$\rm50\,K$), whereas variations
between the two fitting procedures are significant for
temperatures well above the melting point.

A second, longer plateau can be seen in Fig.\ \ref{fig_Temp_Meas}
at $T\approx\rm1100\,K$. We attribute this plateau to the melting
of Ar, the pressure medium. We observed this second plateau in all
measurements in which the sample was heated to the Ar melting
temperature and it occurred at temperatures consistent with the
melting curve of Ar.\cite{boehler2001high}
\subsection{Melting detection}\label{Sec1C}
Methods for detection of the onset of melting by observing changes
in an image of the sample utilize either a speckle image during
heating or a direct image after quenching in
temperature.\cite{schaeffer2012high,jeanloz1996melting,yang2012flash}
In the first method, changes in the speckle image are due to
liquid motion once the hotspot is melted,\cite{jeanloz1996melting}
whereas in the second method changes in the image originate from
changes of texture once the sample is melted and
resolidified.\cite{schaeffer2012high,jeanloz1996melting,yang2012flash}
In both methods, the determination of the exact melting
temperature is somewhat subject to personal interpretation. In
order to avoid this subjectivity, we have developed an image
analysis method to quantify changes in the speckle interference
pattern.

For direct visual observation of melting the hotspot was
illuminated with a HeNe laser beam, which generates a speckle
interference pattern on a CMOS camera (Fig.\ \ref{fig_Opt_Meas}).
At each heating beam power, a series of 8 speckle pattern images
was recorded by the CMOS camera. Laser line filter ($\rm633\,nm$) was used to
prevent the heating laser beam and the thermal radiation from
reaching the camera.

\begin{figure}[ht]
\includegraphics[width=\columnwidth]{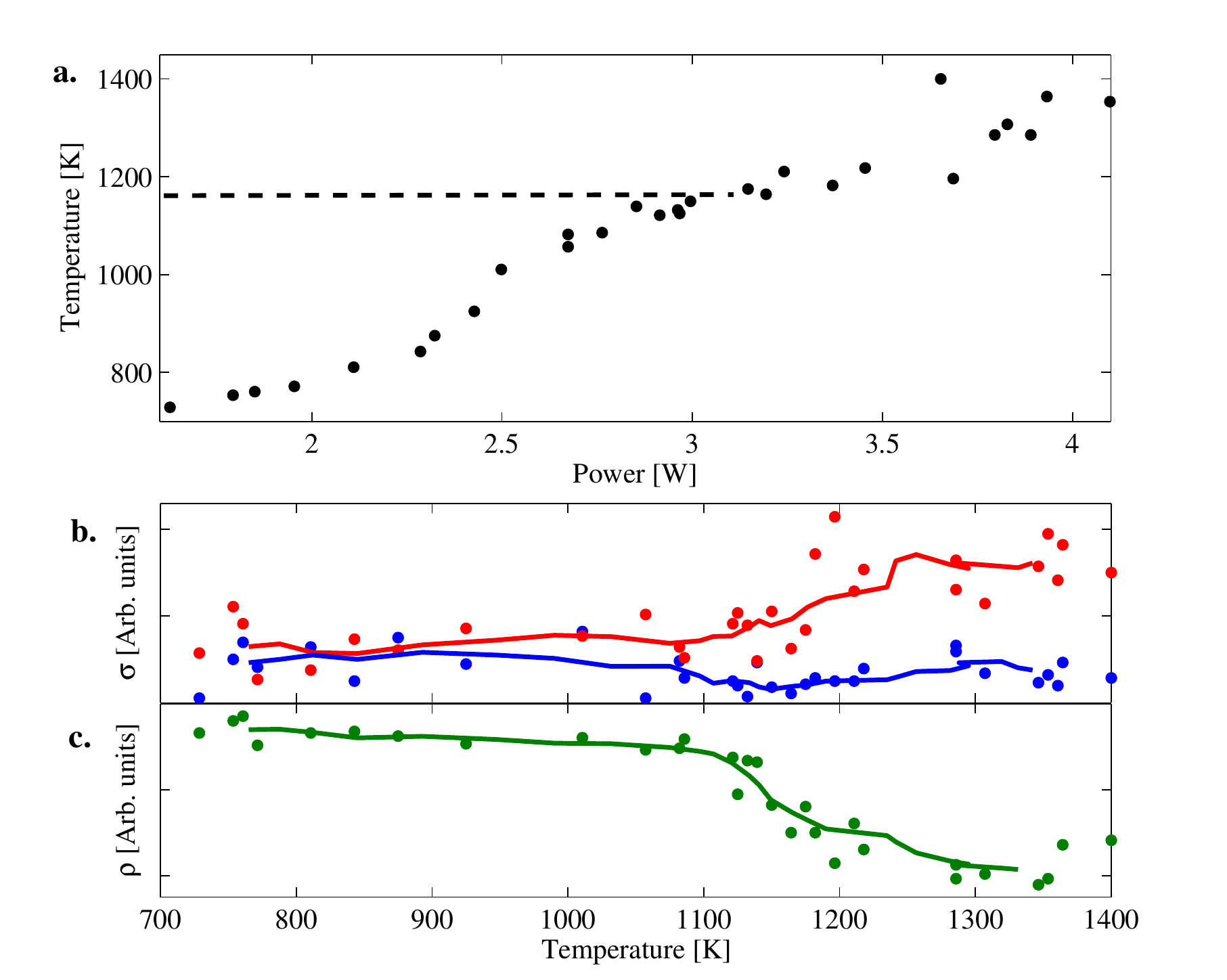}
\caption{ (Color online) Melting point determination. Measurements
were performed on Te sample at a pressure of~$\rm19.5\,GPa$.
 The melting temperature is determined by using three distinct criteria: a.
(upper panel) Observation of changes in the slope of the
temperature as a function of heating beam power. The
dotted line corresponds to the melting temperature b. (middle
panel) An increase in the temporal vibrations of a series of
speckle images taken at each temperature (red line and dots) with
respect to temporal vibrations measured at room temperature after
each heating cycle (blue line and dots). c. (lower panel) A sharp
decrease in the correlation of speckle images taken after heating
the sample, with respect to a reference image taken before heating
the sample. Using these criteria the melting temperature in this
measurement was determined to be $\rm1150\,K$ }
\label{fig_Melt_Det}
\end{figure}

For a quantitative analysis of temporal changes in the speckle
interference pattern, the software calculates the standard
deviation (STD), $\sigma^{}_{ij}$, of each pixel $(i,j)$ in the
series of images,
\begin{align}
\label{eq:2}&\sigma^{}_{ij}=\sqrt{\frac{1}{N}\sum^{N}_{k=1}\left(x^{}_{ijk}-\langle
x^{}_{ij}\rangle\right)^{2}},
\end{align}
where $N=8$ is the number of images, $x^{}_{ijk}$ is the intensity
of pixel $(i,j)$ in the $k$th image and $\langle
x^{}_{ij}\rangle=N^{-1}\sum^{N}_{k=1}x^{}_{ijk}$ is the averaged
intensity of that pixel. The STD per pixel is summed over a region
of interest, giving a parameter that quantifies the amount of
temporal changes generated during the heating process. The outcome
was then compared to the corresponding quantity obtained for a
second series of images recorded at room temperature in the
absence of heating laser beam. Typical plot of the temporal
changes (averaged STD per pixel) at the hotspot as a function of
the measured temperature is shown in the middle panel of Fig.\
\ref{fig_Melt_Det}. At low temperatures there is no apparent
change in the STD of the hotspot during heating and after cooling.
Such differences become evident as the heating beam power
increases. The increasing differences are attributed to melting of
the sample, since liquid motion at the hotspot gives rise to rapid
changes in the speckle interference pattern, which in turn lead to
increase in the STD of the series of images recorded during
heating.

For a more accurate determination of the onset of melting, the
speckle interference patterns were analyzed in a second way.
Typically, melting and resoldification of the sample result in
changes of texture of the melted surface. This is reflected by a
change in the speckle interference pattern after melting and
resolidification. We have quantified such changes by measuring the
correlation of the speckle interference pattern recorded at room
temperature after each heating cycle, with respect to a reference
speckle image recorded at the beginning of the experiment, prior
to any heating process. The correlation coefficient, $\rho$, is
defined as
\begin{align}
\label{eq:3}&\rho=\frac{\sum^{}_{i,j}\left(y^{}_{ij}-\overline{y}\right)\left(z^{}_{ij}-\overline{z}\right)}{\sqrt{\left[\sum^{}_{i,j}\left(y^{}_{ij}-\overline{y}\right)^{2}\right]\left[\sum^{}_{i,j}\left(z^{}_{ij}-\overline{z}\right)^{2}\right]}},
\end{align}
where $y^{}_{ij}$ and $\overline{y}$ are the intensity of pixel
$(i,j)$ and the mean pixel value of the image recorded after a
heating cycle, whereas $z^{}_{ij}$ and $\overline{z}$ are the
corresponding quantities of the reference image. A typical plot of
this correlation as a function of the measured temperature is
shown in the lower panel of Fig.\ \ref{fig_Melt_Det}. One can
clearly observe a drop in the correlation of the speckle
interference patterns once the sample is melted.

To further corroborate the applicability of these two melting
criteria, we also employed a third criterion by monitoring the
changes in the slope of the temperature as a function of heating
beam power (upper panel of Fig.\ \ref{fig_Melt_Det}). This
provides us with three independent criteria for the detection of
melting, as shown in Fig.\ \ref{fig_Melt_Det}.
\section{melting curve of Tellurium}
\label{Sec2} The melting curve of Te was previously studied in a
large volume press up to $\rm10\,GPa$, by means of differential
thermal analysis,\cite{klement1966melting} thermobaric analysis
and electrical resistance measurements.\cite{brazhkin1997high}
However, to the best of our knowledge, no measurements of the
melting curve of Te using LHDAC have been reported to date.
Typically, LHDAC setups are designed for measurements of melting
temperatures above~$\rm1000\,K$, utilizing appropriate optical
components for thermal radiation in the visible range.
Measurements of melting temperatures below~$\rm1000\,K$ require
different optical components sensitive to thermal radiation in the
IR range. Our IR pyrometer, described in detail in Ref.\
\onlinecite{shuker2008ir}, was designed to extend the
applicability of LHDAC experiments to temperatures down to
$\rm500\,K$ and is thus suitable for studying the melting curve of
Te at high pressures.

\begin{figure}[ht]
\includegraphics[width=\columnwidth]{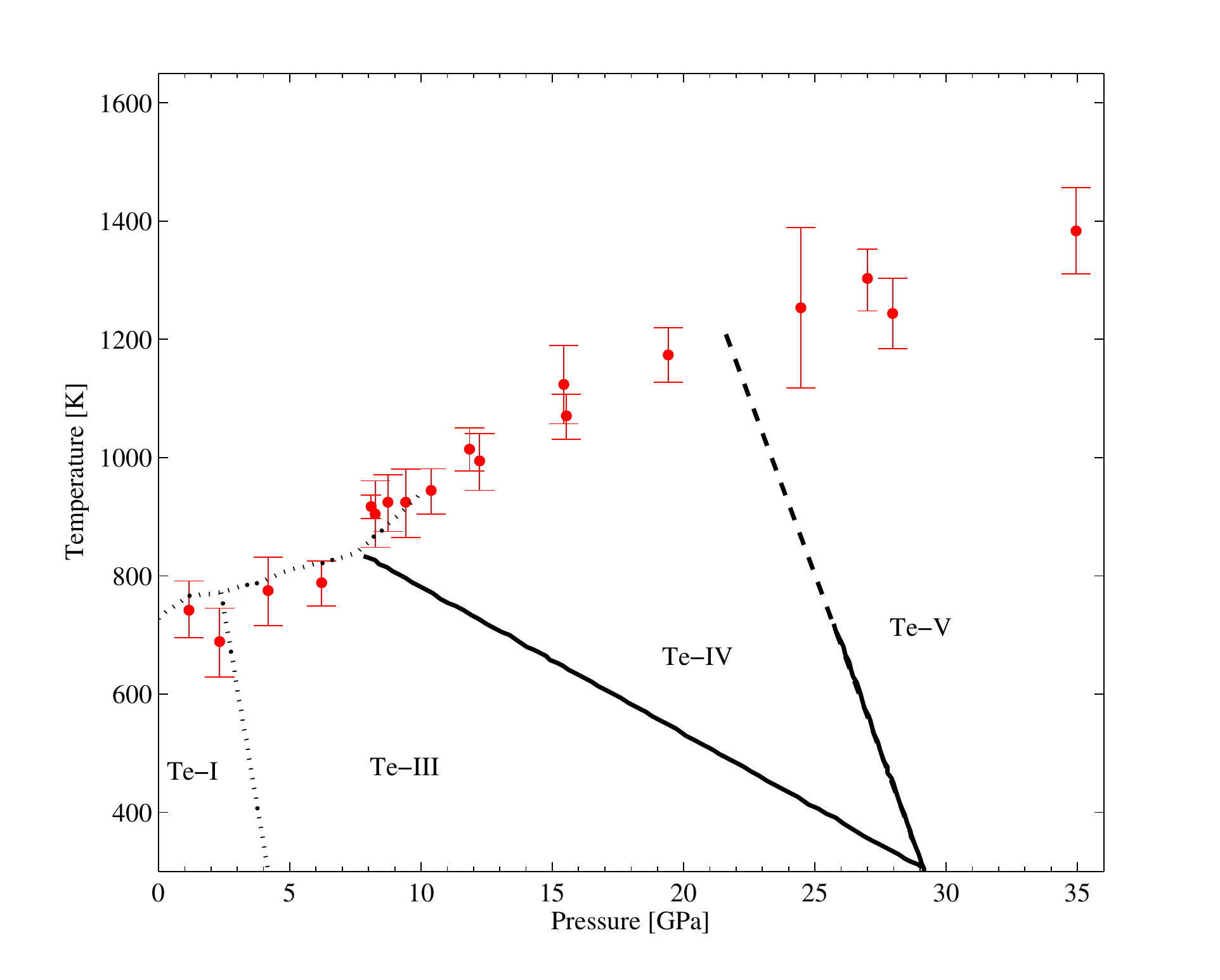}
\caption{(Color online) The phase diagram of Te. Melting points
obtained in this study are denoted by red points. The melting
curve up to~$\rm10\,GPa$ and the I/III phase boundary reported in
Ref.\ \onlinecite{brazhkin1997high} are shown in dotted lines,
whereas the III/IV and IV/V phase boundaries from Ref.~\
\onlinecite{hejny2006phase} are shown in solid
lines.\footnote{Although the transition pressure for the II/III
transition is well determined as~$\rm4.5\,GPa$ at room
temperature, \cite{hejny2003large,hejny2004complex} its
temperature dependence is less determined \cite{hejny2006phase}
and is therefore omitted in the phase diagram plotted in Fig.\
\ref{fig_Melt_Curve}} The extrapolation of the IV/V phase boundary
is plotted in dashed line.} \label{fig_Melt_Curve}
\end{figure}

Using the experimental setup described above, the melting curve of
Te was measured up to a pressure of~$\rm35\,GPa$. At each pressure
we repeated the measurement in at least~$\rm4$ different positions
on the sample surface. At each position, the melting temperature
was determined using at least two of the criteria described in the
previous section. The measured melting curve is shown in Fig.\
\ref{fig_Melt_Curve}. With our improved melting criteria, we could
determine the melting temperature with an estimated error
of~$\rm50\,K$.

Up to~$\rm10\,GPa$ our measurements agree with previous
data.\cite{brazhkin1997high,klement1966melting} Above this point
until~$P\approx\rm25\,GPa$ the melting temperature increases with
increasing pressure but its slope slightly decreases (i.e.,
$d^{2}T^{}_{m}/dP^{2}<0$). At~$P\approx\rm25\,GPa$ a small change
of slope can be observed. This change of slope is possibly
related, via the Clausius-Clapeyron relation, to a solid-solid
phase transition (The IV/V transition, see Fig.\
\ref{fig_Melt_Curve}) reported in Ref.\
\onlinecite{hejny2006phase} at temperatures up to~$\rm700\,K$.
Extrapolation of the IV/V phase boundary (dashed line in Fig.\
\ref{fig_Melt_Curve}) intersects the melting curve
at~$P\approx\rm23\,GPa$, in reasonable agreement with the pressure
at which a change of slope occurs in the melting curve.

\section{Discussion and conclusions}
\label{Sec3} We have demonstrated a simple image analysis
procedure to quantify changes in speckle interference patterns,
which are widely used as a probe for the detection of melting
transitions in LHDAC experiments. Two distinct quantities can be
defined: 1) STD of a series of images recorded during heating and
2) a correlation between an image recorded at room temperature
after each heating cycle and a reference image recorded prior to
heating. Abrupt changes in these two quantities indicate the onset
of liquid motion at the hotspot and changes of texture due to
melting and resolidification, respectively. This method improves
the reliability of melting criteria which are based on changes in
an image of the hotsopt surface.

The image analysis procedure was employed to measure the melting
curve of Te up to~$\rm35\,GPa$. Our measurements are in excellent
agreement with previous measurement of the melting curve of Te
using large-volume press up
to~$\rm10\,GPa$.\cite{brazhkin1997high} At higher pressures our
measurements are in line with the solid-solid coexistence curve,
reported in Ref.\ ~\onlinecite{hejny2006phase}. However, an
additional study of the melting curve in this region is necessary
in order to draw further conclusions. While the use of an image
analysis to improve the accuracy of melting detection was
demonstrated at relatively low temperatures (below 1400K), the
method should also be applicable to temperatures beyond the
measured range.

\begin{acknowledgments}
We thank Y. Greenberg for fruitful discussions.
\end{acknowledgments}


\end{document}